\documentclass[useAMS,usenatbib]{mn2e}

\usepackage{natbib}
\usepackage{amsmath,amsfonts,amssymb}
\usepackage[dvips]{graphicx}
\usepackage{longtable}

\title[Spectral properties of \textit{WMAP} point sources]
  {Follow-up observations at 16 and 33\:GHz of extragalactic sources from \textit{WMAP} 3-year data: I -- Spectral properties\thanks{We request that any reference to this paper cites `AMI Consortium: Davies et al. 2009'}}
\author[AMI Consortium: Davies et al.]
  {AMI Consortium:
  Matthew~L.~Davies,$^1$\thanks{Email: m.davies@mrao.cam.ac.uk}
  Thomas~M.~O.~Franzen,$^1$\thanks{Email: t.franzen@mrao.cam.ac.uk}
  \newauthor
  Rod~D.~Davies,$^2$
  Richard~J.~Davis,$^2$
  Farhan~Feroz,$^1$
  \newauthor
  Ricardo~G\'{e}nova-Santos,$^{1,3}$
  Keith~J.~B.~Grainge,$^{1,4}$
  David~A.~Green,$^1$
  \newauthor
  Michael~P.~Hobson,$^1$
  Natasha~Hurley-Walker,$^1$
  Anthony~N.~Lasenby,$^{1,4}$
  \newauthor
  Marcos~L\'{o}pez-Caniego,$^1$
  Malak~Olamaie,$^1$
  Carmen~P.~Padilla-Torres,$^3$
  \newauthor
  Guy~G.~Pooley,$^1$
  Rafael~Rebolo,$^3$
  Carmen Rodr\'{i}guez-Gonz\'{a}lvez,$^1$
  \newauthor
  Richard~D.~E.~Saunders,$^{1,4}$
  Anna~M.~M.~Scaife,$^1$
  Paul~F.~Scott,$^1$
  \newauthor
  Timothy~W.~Shimwell,$^1$
  David~J.~Titterington,$^1$
  Elizabeth~M.~Waldram,$^1$
  \newauthor
  Robert~A.~Watson$^2$
  and Jonathan~T.~L.~Zwart$^1$\\
  $^1$Astrophysics Group, Cavendish Laboratory,
      19 J.~J.~Thomson Avenue, Cambridge CB3 0HE \\
  $^2$Jodrell Bank Centre for Astrophysics,
      University of Manchester, Manchester M13 9PL\\
  $^3$Instituto de Astrof\'{i}sica de Canarias,
      38200 La Laguna, Tenerife, Canary Islands,
      Spain\\
  $^4$Kavli Institute for Cosmology Cambridge,
      Madingley Road, Cambridge, CB3 0HA}
\date{Accepted ????. Received ????}
\pagerange{\pageref{firstpage}--\pageref{lastpage}}
\pubyear{2009}

\begin{document}
\maketitle
\label{firstpage}
\begin{abstract}
We present follow-up observations of 97 point sources from the \textit{Wilkinson Microwave Anisotropy Probe} (\textit{WMAP}) 3-year data, contained within the New Extragalactic \textit{WMAP} Point Source (NEWPS) catalogue between $-4\degr \leq \delta \leq 60\degr$; the sources form a flux-density-limited sample complete to 1.1\:Jy ($\approx 5\sigma$) at 33\:GHz. Our observations were made at 16\:GHz using the Arcminute Microkelvin Imager (AMI) and at 33\:GHz with the Very Small Array (VSA).

94 of the sources have reliable, simultaneous -- typically a few minutes apart -- observations with both telescopes.  The spectra between 13.9 and 33.75~GHz are very different from those of bright sources at low frequency: 44 per cent have rising spectra ($\alpha_{13.9}^{33.75} < 0.0$), where $S\propto \nu^{-\alpha}$, and 93~per~cent have spectra with $\alpha_{13.9}^{33.75} < 0.5$; the median spectral index is 0.04.

For the brighter sources, the agreement between VSA and \textit{WMAP} 33-GHz flux densities averaged over sources is very good.  However, for the fainter sources, the VSA tends to measure lower values for the flux densities than \textit{WMAP}.  We suggest that the main cause of this effect is Eddington bias arising from variability.

\end{abstract}
\nokeywords

%%%%%%%%%%%%%%%%%%%%%%%%%%%%%%%%%%%%%%%%%%%%%%%%%%%%%%%%%%%%%%%%%%%
\section{Introduction}\label{Introduction}
%%%%%%%%%%%%%%%%%%%%%%%%%%%%%%%%%%%%%%%%%%%%%%%%%%%%%%%%%%%%%%%%%%%

Extragalactic point sources contaminate maps of the cosmic microwave background
(CMB), such as those produced using data from the \textit{Wilkinson Microwave Anisotropy Probe}
(\textit{WMAP}) \citep{bennett}, particularly at frequencies $\lesssim 60$~GHz.  Catalogues of contaminating point sources are,
therefore, a natural by-product of CMB surveys and are an invaluable resource
in the study of point sources at high radio frequencies, the statistical properties of which remain relatively unknown.

The \textit{WMAP} mission has produced all-sky maps of the CMB in five frequency bands between 23 and 94\:GHz.  The point-source catalogue resulting from the original analysis of the 3-year data contains 323 entries \citep{hinshaw};  spectral data from \textit{WMAP} suggest that a large proportion of the detected sources have flat spectra, implying that they belong to a population of compact sources \citep[see e.g.][]{toffolatti}.

\citet{lopez-caniego} have re-analysed the 3-year \textit{WMAP} data to produce the
New Extragalactic \textit{WMAP} Point Source (NEWPS) catalogue of extragalactic point
sources, which contains 369 sources detected at $\mathrm{5\sigma}$ in at least
one of the frequency bands.  The number of sources in the NEWPS catalogue has been maximised by using prior knowledge of source positions from low-frequency data. \citet{hinshaw} neglect the deviations, increasing with frequency, of the \textit{WMAP} point spread function from a Gaussian; the NEWPS estimates use the real beam shape at every frequency. Moreover, \citet{lopez-caniego} have applied a method to correct flux density estimates for the Eddington bias, which leads to an overestimate of the flux densities of faint sources \citep[see][and references therein]{wang}.  

\citet{zotti} made some predictions for high-frequency radio surveys of extragalactic sources. In particular, they claim that the dominant population at $\mathrm{S_{30 GHz} > 1~Jy}$ are blazars in which the radio emission is due to a relativistically beamed jet observed end-on.  However, spectral studies of complete samples selected at cm wavelengths are few.

\citet{bolton} made simultaneous VLA observations at 1.4, 4.8, 22 and 43\:GHz of sources selected from the 9C survey \citep{waldram} at 15~GHz, which has a completeness limit of $\approx$~25\:mJy.  \citet{cleary,cleary_erratum} studied a different sample from the 9C survey, this time with a completeness limit of $\approx 20$~mJy at 15~GHz, using the Very Small Array (VSA) source subtractor at 33\:GHz.  \citet{massardi} have made simultaneous observations with the Australia Telescope Compact Array (ATCA) at 4.8 and 8.6\:GHz of a sample of sources, complete to $\approx$~0.5\:Jy at 20\:GHz, with the ATCA. As a next step, we here address the issue of source spectra at cm wavelengths of higher flux density sources.
 
We have carried out observations of sources contained in the NEWPS catalogue using the Arcminute Microkelvin Imager (AMI) at 16~GHz and the Very Small Array (VSA) at 33\:GHz. For the purpose of obtaining spectra, sources were observed simultaneously (typically within a few minutes) with the AMI and VSA. Finally, we compared \textit{WMAP} flux densities with AMI/VSA flux densities, investigating the effects of source variability.

The layout of this paper is as follows. The remainder of this section describes the NEWPS Catalogue and our source sample. In Section \ref{The Telescopes} we describe the AMI and VSA telescopes. The observations and data reduction are explained in Section \ref{Observations and data reduction}. Section \ref{Spectral properties} discusses the spectral properties of our sample. In Section \ref{Comparison of AMI/VSA fluxes with WMAP fluxes}, we compare AMI/VSA flux densities with \textit{WMAP} flux densities. Finally, we summarize our results in Section \ref{Conclusions}. In an accompanying paper, \citet{franzen} (hereafter Paper II), we present results on the flux density variability of the sources.

%%%%%%%%%%%%%%%%%%%%%%%%%%%%%%%%%%%%%%%%%%%%%%%%%%%%%%%%%%%%%%%%%%%
\subsection{The NEWPS Catalogue}
%%%%%%%%%%%%%%%%%%%%%%%%%%%%%%%%%%%%%%%%%%%%%%%%%%%%%%%%%%%%%%%%%%%

The number of sources in the NEWPS catalogue was maximised by exploiting
low-frequency data.  In the first instance, a catalogue was created from all
those sources with $S \geq$ 500\:mJy in data at 4.85\:GHz from the PMN
\citep{griffith} and GB6 \citep{gregory} surveys.  Where there are holes in the
sky coverage of these surveys they were filled using data from the NVSS survey
\citep{condon} at 1.4\:GHz and the SUMSS survey \citep{bock} at 843\:MHz, again using sources with $S \geq$ 500\:mJy.  The
\textit{WMAP} data were then filtered using the second member of the Mexican Hat Wavelet
family \citep{gonzalez-nuevo06} at the positions of all the sources in the source catalogue.  Any sources detected at $\mathrm{\geq 5\sigma}$, in any one of the frequency channels, were included in the NEWPS 5-$\sigma$ catalogue.

A small number of sources detected in the original \textit{WMAP} analysis were not found in the input
catalogue.  This is because they have strongly inverted spectra, with flux densities $<$
500\:mJy in the low-frequency data.  The \textit{WMAP} data were also filtered at
the positions of these sources and, if detected at $\geq 5\sigma$, were added
to the catalogue.  In total, at 33\:GHz, 224 sources were detected at
$\mathrm{\geq 5\sigma}$, five of which did not appear in the input catalogue.

%%%%%%%%%%%%%%%%%%%%%%%%%%%%%%%%%%%%%%%%%%%%%%%%%%%%%%%%%%%%%%%%%%%
\subsection{Source sample}
%%%%%%%%%%%%%%%%%%%%%%%%%%%%%%%%%%%%%%%%%%%%%%%%%%%%%%%%%%%%%%%%%%%

We selected sources in the NEWPS catalogue detected at $\mathrm{\geq 5\sigma}$ at 33\:GHz and with $-4\degr \leq \delta \leq 60\degr$. These criteria were met by a total of 99 sources. It was not possible to obtain a good observation of J0528+2133, which happens to lie just $1\fdg{5}$ from Tau A, with the VSA. We also excluded J2153+4716, which was identified as an HII region \citep{kuchar}. Having removed these two sources, we were left with a sample containing 97 sources and complete to 1.1\:Jy at 33\:GHz.

%%%%%%%%%%%%%%%%%%%%%%%%%%%%%%%%%%%%%%%%%%%%%%%%%%%%%%%%%%%%%%%%%%%
\section{The Telescopes}\label{The Telescopes}
%%%%%%%%%%%%%%%%%%%%%%%%%%%%%%%%%%%%%%%%%%%%%%%%%%%%%%%%%%%%%%%%%%%

%%%%%%%%%%%%%%%%%%%%%%%%%%%%%%%%%%%%%%%%%%%%%%%%%%%%%%%%%%%%%%%%%%%
\subsection{AMI}\label{The Arcminute Microkelvin Imager}
%%%%%%%%%%%%%%%%%%%%%%%%%%%%%%%%%%%%%%%%%%%%%%%%%%%%%%%%%%%%%%%%%%%

The AMI \citep[see AMI Consortium:][for a full description of the instrument]{zwart} consists of two separate telescopes, the Large (AMI-LA) and Small (AMI-SA) Arrays; all the AMI data presented in this paper were obtained from the AMI-SA. The AMI-SA operates at frequencies between 13.9 and 18.2\:GHz with the passband divided into six channels of 0.72-GHz bandwidth. The average frequencies of the 6 usable bands are given in Table~\ref{tab:fluxcal} (the AMI-SA has two frequency channels below 13.9~GHz which are not routinely used, owing to interference from satellites). The centre frequency is $\approx 16.1$~GHz.  The results presented in this paper are a combination of continuum and channel flux densities.  We make it clear in the results sections which of these we are using.

The primary beam of the telescope at 16~GHz is $\approx 20$~arcmin FWHM.  The synthesised beam, which is an effective measure of the resolution of the telescope, varies with frequency channel and observation declination.  However, it is typically $\approx 3$~arcmin FWHM.  The telescope measures a single, linear polarisation: Stokes parameter $\mathrm{I + Q}$. It has a flux sensitivity of $\approx 30$~mJy in one second and is able to observe at declinations $> -15\degr$.

\begin{table}
 \caption{Assumed flux densities for sources used for AMI flux density calibration. Note that the individual channel errors are not independent and that the error on each calibration error is about half a per cent.}
 \label{tab:fluxcal}
 \begin{tabular}{@{}c c c c c }
 \hline
 Channel & $\bar{\nu}$/GHz & \multicolumn{2}{c}{$S$/Jy} & R.M.S. calibration \\
         &                 & 3C286 & 3C48             &   error/per~cent \\
 \hline
 3&14.2&3.61&1.73&6.5 \\
 4&15.0&3.49&1.65&5 \\
 5&15.7&3.37&1.57&4 \\
 6&16.4&3.26&1.49&3.5 \\
 7&17.1&3.16&1.43&4 \\
 8&17.9&3.06&1.37&7 \\
 \hline
 \end{tabular}
\end{table}

%%%%%%%%%%%%%%%%%%%%%%%%%%%%%%%%%%%%%%%%%%%%%%%%%%%%%%%%%%%%%%%%%%%
\subsection{VSA}
%%%%%%%%%%%%%%%%%%%%%%%%%%%%%%%%%%%%%%%%%%%%%%%%%%%%%%%%%%%%%%%%%%%

The main array of the VSA \citep[see][for a detailed description of the instrument]{watson}, which was used to obtain the VSA data presented in this paper, has a single frequency channel centred at 33\:GHz with 1.5\:GHz bandwidth.  The VSA has several observing configurations.  The observations presented here were made with the VSA in its super-extended configuration \citep[see][]{genova-santos}. 

The flux sensitivity of the telescope in the super-extended configuration is $\approx 2.7$~Jy in one second.  The primary beam is 72~arcmin FWHM.  Again, the synthesised beam varies with observation declination.  A typical value, however, is 7~arcmin FWHM.  The VSA can observe in the declination range $-4\degr \leq \delta \leq 60\degr$.  The telescope is not equatorially mounted, so the linear polarisation measured changes with the hour angle of observation.  At a parallactic angle of $0\degr$ the telescope measures Stokes parameter $\mathrm{I - Q}$.

%%%%%%%%%%%%%%%%%%%%%%%%%%%%%%%%%%%%%%%%%%%%%%%%%%%%%%%%%%%%%%%%%%%
\section{Observations and data reduction}\label{Observations and data reduction}
%%%%%%%%%%%%%%%%%%%%%%%%%%%%%%%%%%%%%%%%%%%%%%%%%%%%%%%%%%%%%%%%%%%

%%%%%%%%%%%%%%%%%%%%%%%%%%%%%%%%%%%%%%%%%%%%%%%%%%%%%%%%%%%%%%%%%%%
\subsection{AMI}\label{AMI Observations and data reduction}
%%%%%%%%%%%%%%%%%%%%%%%%%%%%%%%%%%%%%%%%%%%%%%%%%%%%%%%%%%%%%%%%%%%

Observations of the \textit{WMAP} sources were made using the AMI-SA during fourteen separate observing runs between 2007 April and 2008 September.  The majority of the sources were observed, at irregular intervals, at least three times during this period.  During the earlier observing runs, which typically lasted about 48 hours, the sources were observed twice (at different hour angles) for ten minutes, although some of the fainter sources were observed for longer to match the observing time on the VSA.  This two-observation scheme provided a useful check of the reliability of our calibration procedures (see below).  The two datasets were then concatenated so as to improve the signal-to-noise and the \textsc{uv} coverage of the aperture plane.  Having  established the reliability of our calibration procedures, during the later observing runs the sources were observed only once, typically for twenty minutes each.

Data reduction was carried out using \textsc{reduce}, our software developed, originally, for the VSA and later modified for the AMI.  This applies path delay corrections, automatic flags for intereference, pointing errors, shadowing and hardware faults, applies phase and amplitude calibrations, Fourier transforms the data into the frequency domain, and writes it out in \textsc{uv} \textsc{fits} format for imaging in the \textsc{aips}\footnote{http://www.aips.nrao.edu} package.

Flux density calibration was performed using short observations of 3C48 and 3C286 interspersed with the observations of the \textit{WMAP} point sources.  The flux densities assumed for these sources, used for calibrating each of the frequency channels, are consistent with those of \citet{baars}.  Since we measure a different polarisation from that measured by \citeauthor{baars} ($\mathrm{I + Q}$, as opposed to I), we correct for the polarisations of the calibrator sources by interpolating VLA data collected at 5, 8 and 22~GHz.  The assumed flux densities in each of the AMI frequency channels are listed in Table~\ref{tab:fluxcal}.

We correct for changing air mass and variations in atmospheric conditions both during the observations and between the observations of the sources and the flux density calibrators by monitoring a modulated noise signal, injected at the front-end of each antenna. The data are also weighted by comparing the ratio of the power of the modulated noise signal to the total power input to the correlator (which is kept constant), to that obtained in cool, dry, clear weather conditions.  Samples are flagged if the data were taken in poor weather conditions, or if there were large changes in the weather between the observation of the source and the flux density calibrator.

The telescope is not expected to be phase stable over the duration of an entire observing run.  It is, however, phase stable over the length of an individual observation and, since the SNRs for all sources were sufficiently large, we were able to use them as their own phase calibrators for the purpose of absolute phase calibration.  This negated the requirement for observations of interleaved phase calibrators.  We find that implementing this scheme for phase calibration produces a significant decrease in the dispersion of the flux densities measured for the cross-calibrated observations of 3C286 and 3C48.  It also reduces the scatter in the two flux densities, measured for each of the \textit{WMAP} point sources, within the same observing run.

The \textsc{uv} \textsc{fits} data output from \textsc{reduce} were imaged using the \textsc{aips} package.  Maps were produced for channels 3 to 8 individually and combined.  The great majority of the sources were not resolved by the AMI-SA.  In these cases, the quoted flux density is the peak value from the dirty map. For the few sources that were resolved we first \textsc{clean} the map and then measure the integrated flux density.  Following a similar method to \citet{rees}, we sum contiguous pixels down to a contour level of half the peak flux density value to give $I_{mp}$. We then apply the same operation to the beam with a contour level of half the height of the maximum on the beam, to give $I_{bm}$. The integrated flux density is then taken to be $I_{mp}$/$I_{bm}$.  

\subsubsection{Flux density error estimation}

We checked the flux density calibration of the telescope by comparing the two observations of each source from individual observing runs, making the assumption that the variability timescale was long compared to the duration of the observing run.  We also cross-calibrated observations of our flux density calibrators.  Both tests indicated that the flux density calibration of the telescope is consistent to $\mathrm{\approx~3}$ per cent r.m.s.

We have taken a conservative approach and assumed that the error, $\sigma$, on a measured continuum flux density, $S$, consists of a 4 per cent calibration error added in quadrature with the thermal noise on the map, $n$ (which is measured far from the primary beam). We add the calibration error in quadrature with the thermal noise because these two errors are uncorrelated.  The total error is therefore given by $\sigma = \sqrt{(0.04S)^2 + n^2}$.  Similar tests were performed to measure the calibration errors on the individual frequency channels, which are shown in Table~\ref{tab:fluxcal}.  Errors quoted on channel flux densities consist of the map noise added in quadrature with the calibration error for the relevant channel.

%%%%%%%%%%%%%%%%%%%%%%%%%%%%%%%%%%%%%%%%%%%%%%%%%%%%%%%%%%%%%%%%%%%
\subsection{VSA}
%%%%%%%%%%%%%%%%%%%%%%%%%%%%%%%%%%%%%%%%%%%%%%%%%%%%%%%%%%%%%%%%%%%

Observations of the \textit{WMAP} sources were performed during ten separate observing runs between 2007 February and 2008 July. The majority of the sources were observed, at irregular intervals, at least four times during this period. Observations were 10--20 minutes in length. During most observing runs, which typically spanned 24--48 hours, each of the sources was observed twice. A small proportion of observations was discarded for reasons unrelated to source characteristics, such as interference from the Sun, Moon or planets. 

A full description of the VSA calibration process is presented in \cite{watson}.  Briefly, the absolute flux density calibration of VSA data is determined from observations of Jupiter, the brightness temperature ($T_{Jup} = 146.6 \pm 0.7$~K) of which is taken from the \textit{WMAP} 5-year data.  These, in turn, were calibrated on the CMB dipole \citep{hill}. This flux density scale is transferred to our other calibrator sources: Tau A, Cas A, Cyg A and Saturn.  The specifications for the super-extended configuration, such as the correction for the fact that Tau A and Cas A are partially resolved in the longest baselines, are essentially the same as those adopted for the extended configuration and are explained in \cite{dickinson}. The assumed flux density of Cyg A, which is unresolved by the VSA, is 36.7\:Jy at 33~GHz.

Data reduction was carried out using the \textsc{reduce} package written specifically for the VSA (see Section~\ref{AMI Observations and data reduction}).  Amplitude and phase calibrations were performed using observations of Cas A, Cyg A, Saturn and Tau A interspersed with the observations of the \textit{WMAP} sources.  A calibration scheme addressing temperature effects on the phase was adopted \citep[see][]{lancaster}. The data were corrected for changes in system temperature and in atmospheric opacity with elevation, based on the monitoring of modulated noise signals injected into the VSA system. The correction was typically a few per cent. In order to achieve the optimum overall noise level, the data were re-weighted based on the r.m.s. noise of each baseline.

\subsubsection{Phase calibration}

The \textsc{uv} \textsc{fits} data output from \textsc{reduce} were imaged using the \textsc{aips} package.  To correct phase errors, we self-calibrated datasets making the assumption that all the sources are unresolved by the VSA. To achieve this, we employed the \textsc{calib} task using a point-source model.  We used the phase-only solution mode, with a solution interval encompassing the entire observation.  We ran the task on each dataset but only applied solutions to the data if a solution was found for \textit{each} antenna.  We found this to be the case for a total of 176 observations ($\approx 25$~per~cent of observations), which typically had signal-to-noise ratios (SNRs) $\gtrsim 15$. The remaining observations had insufficient SNR for self-calibration to be applied successfully (we explain how we addressed this problem in the next paragraph). We then produced dirty maps using the \textsc{imagr} task, applying \textit{natural} weighting in order to maximise the SNR. In each map, we extracted the peak flux density inside a square at the map centre with half-width of 10~arcmin.

\begin{figure}
\includegraphics[scale=0.6,angle=0]{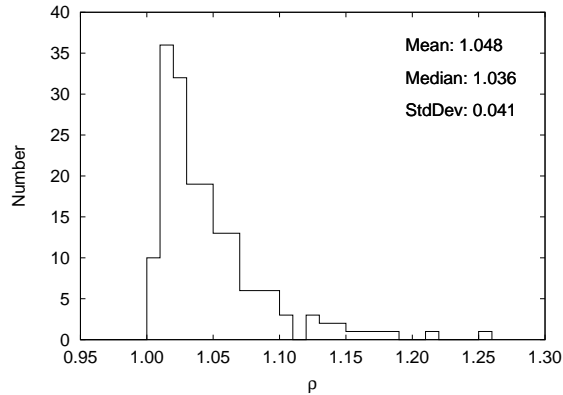}
\caption{The distribution of the ratios, $\rho$, of the flux densities with self-calibration to those without for the 176 observations with sufficient SNR for the \textsc{calib} task to be applied successfully.}
\label{fig:freq}
\end{figure}

For the 176 observations with sufficient SNR for the \textsc{calib} task to be applied successfully, we also produced maps using the un-self-calibrated data in order to compare flux densities with and without self-calibration. Fig. \ref{fig:freq} shows the distribution of the ratios, $\rho$, of the flux densities with self-calibration to those without. We emphasize that we are dealing with peak flux densities. As expected, phase errors systematically result in flux densities being underestimated. The mean and median values of $\rho$ are 1.048 and 1.036 respectively. In response to these findings, for the remaining, un-self-calibrated observations, we, accordingly, have multiplied flux densities by 1.048 as a correction. 

\subsubsection{Flux density error estimation}

The thermal noise was measured from the time-series dispersion of the visibilities in \textsc{reduce} and is, for a typical observation, 0.15--0.20\:Jy. We checked the flux density calibration of the telescope by cross-calibrating observations of our flux density calibrators. This test indicated that, from the amplitude alone, the flux density calibration of the telescope is subject to an error of $\mathrm{\approx~4}$˜ per cent r.m.s. From the standard deviation of $\rho$ (0.041), we deduced that, from the phase alone, the flux density calibration is subject to an error of $\mathrm{\approx~4}$ per cent r.m.s. The fact that this error has a non-Gaussian distribution, as is clearly apparent in Fig.~\ref{fig:freq}, should not be a significant problem, because, in the cases where it is not possible to apply self-calibration, the total error is dominated by the thermal noise.

We have, therefore, assumed that, for each source sufficiently bright to be used as its own phase calibrator, the total error on a measured flux density, $S$, is given by $\sigma = \sqrt{(0.04S)^2 + n^2}$, where $n$ is the thermal noise.  Otherwise, taking a conservative approach, the total error is taken to be $\sigma = \sqrt{(0.08S)^2 + n^2}$.  We add the calibration error in quadrature with the thermal noise because these two errors are uncorrelated.
 
We checked our error estimates by comparing the two observations of each source from individual observing runs, making the assumption that the variability timescale was long compared to the duration of the observing run. From a total of 687 observations, there were 248 such pairs of observations. A $\mathrm{\chi^{2}}$-test for the difference between the two flux densities produced a reduced $\mathrm{\chi^{2}}$-value of 0.98 for 247 degrees of freedom.  The probability of exceeding this value, by chance, is 0.58.  As a result, we are confident in our error estimates. 

For each source observed twice during the same observing run, we have treated the two observations as a single observation. If self-calibration was applied to both observations, the two datasets were concatenated before imaging and the quoted flux density, $S$, is the peak flux density in the resulting map. Since, for a self-calibrated source, the phase errors have been corrected, the quoted error is given by $\sigma = \sqrt{(0.04S)^2 + n^2}$, where $n = \left(1/n^2_{A} + 1/n^2_{B}\right)^{-1/2}$, and $n_{A}$ and $n_{B}$ are the thermal noises in the two individual observations.

Because of possible discrepancies in source positions arising from phase errors, we concatenated the two datasets only in cases where it was possible to apply self-calibration to both observations. Otherwise, the quoted flux density is given by
\begin{eqnarray}
    S = \frac {S_{A}/n^2_{A} + S_{B}/n^2_{B}}{1/n^2_{A} + 1/n^2_{B}}\mathrm{,}
\end{eqnarray}
where $S_{A}$ and $S_{B}$ are the two measured flux densities with thermal noises $n_{A}$ and $n_{B}$ respectively.

If self-calibration was applied to neither of the two observations, the quoted error is given by $\sigma = \sqrt{(0.08S)^2 + n^2}$ and in, the few cases where it was applied to one of the two observations, $\sigma = \sqrt{(0.06S)^2 + n^2}$. Note that we have considered the worst case scenario where the flux density calibration errors in the two individual observations are correlated (which is a possibility given how the observations were scheduled). 

\begin{figure}
\includegraphics[scale=0.6,angle=0]{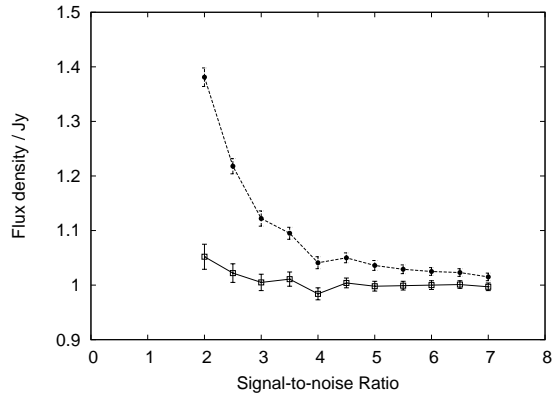}
\caption{Results of a simulation to investigate the bias resulting from measuring peak flux densities at low SNR with the VSA.  We simulated pointed observations of a 1.0-Jy point source with different SNRs, performing 500 realisations per SNR.  Open squares show the average flux densities measured at the pixel corresponding to the map centre; filled circles show the average peak flux density inside a square, centred at the map centre, with half-width of 10~arcmin.}
\label{fig:noise_bias}
\end{figure}

\subsubsection{Peak-flux-density bias}

We have identified a bias in our measurements at low SNR, which results from measuring peak flux densities as opposed to those at the exact source positions. Low-frequency identifications are given for each of the sources in the NEWPS catalogue.  For increased accuracy, we used the coordinates of sources listed as likely matches in the 4.85-GHz PMN or GB6 catalogues.  Because of the VSA's very different observing frequency, we checked for positional differences (these were typically 1.5~arcmin), but, in fact, these were mainly caused by phase calibration errors in the VSA data.  Therefore, in each case, rather than measuring the flux density at the pixel corresponding to the pointing centre, we measured the peak flux density inside a square at the map centre with half-width of 10~arcmin (i.e. within $\approx$ one synthesised beam from the pointing centre). However, at low SNR, the peak flux densities are biased to be slightly high, because of the increased likelihood of peak positions lying on top of positive noise fluctuations.

We have perfomed simulations to investigate this bias, the results of which are shown in Fig.~\ref{fig:noise_bias}. Pointed observations of a 1.0-Jy point source with SNRs ranging between 2.0 and 7.0 were simulated. We used a typical \textsc{uv} coverage for the VSA and produced dirty maps using the \textsc{aips} package. We performed a total of 500 realisations per SNR. For each simulated map, we measured the flux density at the pixel corresponding to the map centre and the peak flux density inside a square at the map centre with half-width of 10~arcmin.

As expected, we found that there is no bias if flux densities are measured at the correct positions, no matter how low the SNR. We found that, at SNR = 5.0, peak flux densities are on average $\mathrm{\approx~4}$ per cent higher than the true flux density.  The effect becomes more severe as the SNR decreases further, but even at SNR = 3.0 (only $\approx 3$~per~cent of observations are below this SNR), peak flux densities are on average only $\approx 12$~per~cent higher than the true flux densities.  Since the magnitude of the bias is small relative to the thermal noise, we do not expect it to have any significant effect on eventual results and have, therefore, not corrected the VSA flux densities for this effect.

%%%%%%%%%%%%%%%%%%%%%%%%%%%%%%%%%%%%%%%%%%%%%%%%%%%%%%%%%%%%%%%%%%%
\section{13.9 to 33.75~GHz Spectral properties}\label{Spectral properties}
%%%%%%%%%%%%%%%%%%%%%%%%%%%%%%%%%%%%%%%%%%%%%%%%%%%%%%%%%%%%%%%%%%%

\begin{figure}
\includegraphics[scale=0.35,angle=270]{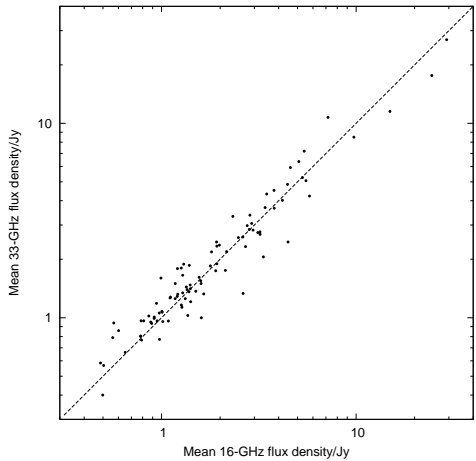}
\caption{Mean flux density at 16 versus that at 33\:GHz.  The dashed line corresponds to $\alpha_{16}^{33} = 0$}
\label{fig:mean_ami_flux_vs_mean_vsa_flux}
\end{figure}

In Fig.~\ref{fig:mean_ami_flux_vs_mean_vsa_flux} we have plotted the mean flux density at 16~GHz versus that at 33~GHz for each of the sources in our sample.  The plot demonstrates that the sample contains a large proportion of sources with rising spectra.  The mean AMI and VSA flux densties, along with the number of observations, are given for each source, in Table~\ref{tab:sources}.

We have simultaneous observations of most of the sources in the sample, a significant advantage over other work.  The obervations on the two telescopes were typically performed within a few minutes of one another; at worst they were separated by two days.  We have at least a single pair of reliable, simultaneous observations for 94 of the sources.  For the remaining three sources in our sample we have no reliable simultaneous observations; this is due to hardware problems or interference during the observations.

We have fitted power-law spectra to these simultaneous data, defining spectral index, $\alpha$, by $S\propto \nu^{-\alpha}$.  In fitting the spectra we have made use of the spectral data from the six useable AMI frequency channels and the single VSA channel.  The frequencies at the lower end of the AMI and the upper end of the VSA passbands are 13.9 and 33.75~GHz respectively.  We, therefore, quote spectra, $\alpha_{13.9}^{33.75}$, between these two frequencies.

The spectra were fitted by sampling using a Markov Chain Monte Carlo technique from a uniform likelihood function.  This method takes into account the asymmetric errors in $\ln~S$ and enables the calculation of an error estimate for each spectral index, taking account of the posterior distribution.  We find that, if there were to be a 5~per~cent difference in the flux density scales between the two telescopes, the resulting systematic error in the spectral index would be 0.07.

\begin{figure}
\includegraphics[scale=0.35,angle=270,bb=0.7in -0.5in 8in 8.5in,clip=]{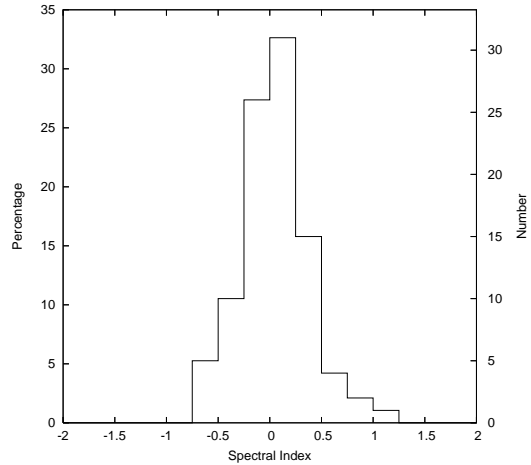}
\caption{Histogram of the spectral index distribution $\alpha_{13.9}^{33.75}$.}
\label{fig:histogram}
\end{figure}

\begin{figure*}
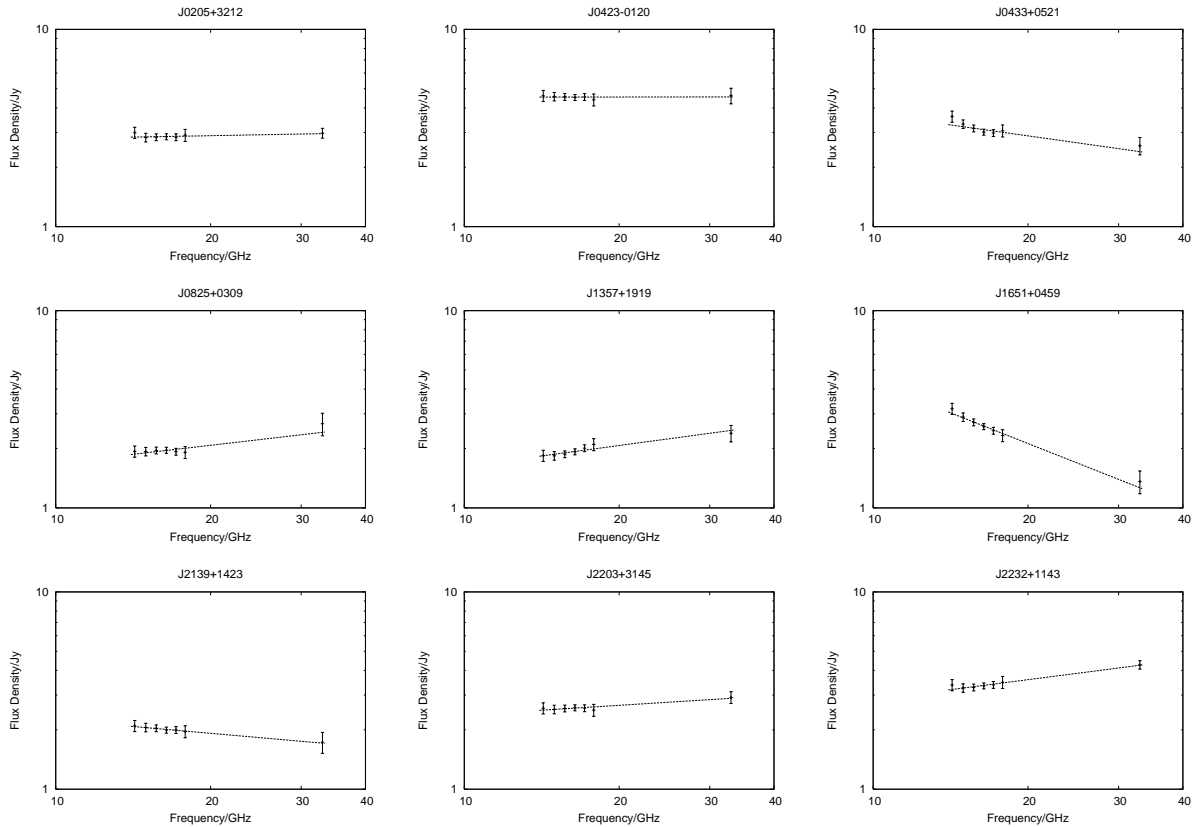

  \includegraphics[angle=270,scale=0.20,bb=0.7in 0.5in 8in 11in,clip=]{figures/RUN_02_J0205+3212_BOTH.epsi}
  \includegraphics[angle=270,scale=0.20,bb=0.7in 0.5in 8in 11in,clip=]{figures/RUN_02_J0423-0120_BOTH.epsi}
  \includegraphics[angle=270,scale=0.20,bb=0.7in 0.5in 8in 11in,clip=]{figures/RUN_02_J0433+0521_BOTH.epsi} \\
  \includegraphics[angle=270,scale=0.20,bb=0.7in 0.5in 8in 11in,clip=]{figures/RUN_02_J0825+0309_BOTH.epsi}
  \includegraphics[angle=270,scale=0.20,bb=0.7in 0.5in 8in 11in,clip=]{figures/RUN_02_J1357+1919_BOTH.epsi}
  \includegraphics[angle=270,scale=0.20,bb=0.7in 0.5in 8in 11in,clip=]{figures/RUN_02_J1651+0459_BOTH.epsi} \\
  \includegraphics[angle=270,scale=0.20,bb=0.7in 0.5in 8in 11in,clip=]{figures/RUN_02_J2139+1423_BOTH.epsi}
  \includegraphics[angle=270,scale=0.20,bb=0.7in 0.5in 8in 11in,clip=]{figures/RUN_03_J2203+3145_BOTH.epsi}
  \includegraphics[angle=270,scale=0.20,bb=0.7in 0.5in 8in 11in,clip=]{figures/RUN_02_J2232+1143_BOTH.epsi} \\
    \caption{Examples of spectra between 13.9 and 33.75~GHz. Note that the errors on the AMI flux densities measured in different frequency channels are not independent.}
  \label{fig:spectrum_sources}
\end{figure*}

We provide a histogram of the spectral index distribution $\alpha_{13.9}^{33.75}$ for the 94 sources in Fig.~\ref{fig:histogram}.  For those sources with multiple pairs of simultaneous observations, we have used the mean spectral index in plotting the histogram.  The spectral index, fitted to the simultaneous observations, of each source, along with the associated error, are listed in Table~\ref{tab:sources}.  The table shows that the typical errors are smaller than the bin size used in plotting the histogram.  In Fig.~\ref{fig:spectrum_sources} we have provided spectra for a sample of typical sources.  Of the 94 sources, 41 (44 per cent) have rising spectra ($\alpha_{13.9}^{33.75} < 0$) and 87 (93~per~cent) have spectra with $\alpha_{13.9}^{33.75} < 0.5$.  The median spectral index is 0.04.

\section{Comparison of AMI/VSA and \textit{WMAP} flux densities}\label{Comparison of AMI/VSA fluxes with WMAP fluxes}

In Fig.~\ref{fig:VSA_versus_WMAP}, we compare our 33-GHz flux densities with the 33-GHz flux densities measured by \textit{WMAP} and, in figure \ref{fig:WMAP_versus_AMI}, we compare our 16-GHz continuum flux densities with the 23-GHz flux densities measured by \textit{WMAP}.  Because the sources are variable (see Paper II), we used the mean AMI/VSA flux density for each source.  Since the level of variability for many of the sources is large compared to the errors on the flux density measurements, we have not included error bars in these figures.

\begin{figure}
\includegraphics[scale=0.6,angle=0]{figures/VSA_versus_WMAP.epsi}
\caption{Comparison of VSA 33-GHz flux densities with \textit{WMAP} 3-year 33-GHz flux densities, with a line indicating equal flux density values.}
\label{fig:VSA_versus_WMAP}
%\end{figure}
%\begin{figure}
\includegraphics[scale=0.6,angle=0]{figures/WMAP_versus_AMI.epsi}
\caption{Comparison of AMI 16-GHz flux densities with \textit{WMAP} 3-year 23-GHz flux densities, with a line indicating equal flux density values.  We used flux densities quoted in the NEWPS $3\sigma$ catalogue for the faintest sources that did not appear in the NEWPS 5-$\sigma$ catalogue.  These sources are represented as open squares.}
\label{fig:WMAP_versus_AMI}
\end{figure} 

\begin{figure}
\includegraphics[scale=0.6,angle=0]{figures/bias_VSA_versus_WMAP_LOG.epsi}
\caption{Comparison of the mean VSA 33-GHz flux density with the mean \textit{WMAP} 3-year 33-GHz flux density in three groups of sources, a low, medium and high flux density group, with a line indicating equal flux density values. The error bars are standard errors of the means. Calculations were performed in log space.}
\label{fig:bias_VSA_versus_WMAP}
%\end{figure}
%\begin{figure}
\includegraphics[scale=0.6,angle=0]{figures/bias_AMI_versus_WMAP_LOG.epsi}
\caption{Comparison of the mean AMI 16-GHz flux density with the mean textit{WMAP} 3-year 23-GHz flux density in three groups of sources, a low, medium and high flux density group, with a line indicating equal flux density values. The error bars are standard errors of the means. Calculations were performed in log space.}
\label{fig:bias_AMI_versus_WMAP}
\end{figure}

We have used the Bayes-corrected version \citep[see][]{herranz} of the NEWPS Catalogue to take into account Eddington bias.  Close to the survey completeness limit, source flux densities are biased high. The presence of noise, and the fact that faint sources are more numerous than bright ones, result in an increased likelihood for peak positions to lie on top of positive fluctuations. The correction for each source in the NEWPS catalogue is calculated from the SNR and the slope of the source counts.  The source counts are described by a power law, which is estimated from the NEWPS flux densities themselves.

The flux density calibrations of all the \textit{WMAP} channels have been reassessed through estimates of the effective beam areas and correction factors have been derived \citep{gonzalez-nuevo08}.  These correction factors are 1.05, 1.086, 1.136 and 1.15 at 23, 33, 41 and 61 GHz respectively. We have corrected the NEWPS flux densities by these self-calibration factors.  

We note that eight sources have no flux densities quoted at 23 GHz in the NEWPS 5-$\sigma$ catalogue because they are not detected at $\mathrm{\geq 5\sigma}$.  For these sources, we used the flux densities quoted in the NEWPS 3-$\sigma$ catalogue.  These sources are represented as open squares in Fig.~\ref{fig:WMAP_versus_AMI}.

Figures.~\ref{fig:VSA_versus_WMAP} and \ref{fig:WMAP_versus_AMI} indicate that there is generally very good agreement between the AMI/VSA flux densities and the \textit{WMAP} ones.  Discrepancies between the flux densities are expected given our findings in Paper II concerning the general level of variability of the sources in the sample.  Our observations were carried out in 2007 and 2008, i.e. in a period after that of \textit{WMAP} 3-year maps (obtained by averaging over the data collected during 2001-2004).  We note that the frequencies at which the AMI and \textit{WMAP} results are compared are not the same.  However, we do not expect this to contribute significantly to the scatter in Fig.~\ref{fig:WMAP_versus_AMI} because the two frequencies are relatively close to one another and, as discussed in Section~\ref{Spectral properties}, a large proportion of the sources have flat spectra.

We have checked whether there is a good agreement between the VSA and \textit{WMAP} flux densities for a very bright non-variable source in our sample, Cyg A (J1959+4034).  This source is also one of the VSA calibrator sources, with an assumed flux density of $\mathrm{36.7 \: Jy}$. The overall extent of the source is 2.1~arcmin and it is unresolved by the VSA and \textit{WMAP}.  No variability in the core of Cyg~A has been reported, at any frequency.  In any event, the core represents a small fraction of the total flux density.  \citet{alexander} made observations of Cyg A with the Cambridge 5-km radio telescope and measured the flux density of the central component at 15.4\:GHz to be just $1.22 \pm 0.2$~Jy.  We found that the flux density quoted by \textit{WMAP} $\mathrm{(37.87 \pm 0.49~Jy)}$ is in excellent agreement with the flux density measured, here, by the VSA $\mathrm{(37.43 \pm 0.84~Jy)}$.

\addtocounter{subsubsection}{-3}
\subsubsection{Offset between AMI/VSA and \textit{WMAP} flux densities}

It is also apparent in figs. \ref{fig:VSA_versus_WMAP} and \ref{fig:WMAP_versus_AMI} that there is a systematic offset at faint flux densities, the AMI/VSA flux densities tending to be lower than the \textit{WMAP} ones. We quantified this effect as follows: for each source, we calculated the mean AMI/VSA and \textit{WMAP} flux density and on this basis, split the sources into three equally sized groups, a low-, medium- and high-flux-density group. In each group, we then compared the mean AMI/VSA flux density with the mean \textit{WMAP} flux density. Given the large range of flux densities, we performed calculations in log space. The results are shown in Figs. \ref{fig:bias_VSA_versus_WMAP} and \ref{fig:bias_AMI_versus_WMAP}. The error bars are standard errors of the means.

In the high-flux-density group, the mean AMI/VSA and \textit{WMAP} flux densities are in excellent agreement. However, in the low-flux-density group, the mean AMI and VSA flux densities are significantly lower than the mean \textit{WMAP} flux density, the discrepancies being significant at a level of 3.0$\sigma$ (AMI) and 4.9$\sigma$ (VSA). In the medium-flux-density group, the effects are significant at levels of 4.2$\sigma$ (AMI) and 2.0$\sigma$ (VSA). This statistical analysis suggests that the systematic offset at faint flux densities is very unlikely to be due to chance and requires an explanation. 

One possible issue is that the AMI beam is $\approx 3$~arcmin, the VSA beam $\approx 7$~arcmin, the \textit{WMAP} beam in the K-band $\approx 53$~arcmin and the \textit{WMAP} beam in the Ka-band $\approx 40$~arcmin.  However, incomplete sampling of extended sources with AMI and the VSA cannot explain the bias because, as discussed in Section~\ref{AMI Observations and data reduction}, practically all sources were found to be point-like at the AMI resolution.

Another possible factor is source confusion with \textit{WMAP}.  In order to have a significant effect, this would require more than one source with $S \gtrsim$ 1\:Jy to lie within a \textit{WMAP} resolution element.  This is very unlikely given the number of sources with $S \gtrsim$ 1\:Jy detected by \textit{WMAP} in all-sky maps.  The NEWPS catalogue contains 224 sources at 33\:GHz; there are $\approx 500$~WMAP~beam~areas per source in the Ka-band.

Given the length of time between our observations and the WMAP ones and the general level of variability (see Paper II), we suggest that the overwhelming cause of the systematic effect is Eddington bias arising from variability.

\begin{figure}
\includegraphics[scale=0.6,angle=0]{figures/WMAP_versus_WMAP_selyr1.epsi}
\caption{Comparison of flux densities in year 1 with flux densities in years 2 to 5 for \textit{WMAP} 33-GHz data.  Sources were selected in the first-year map, with a line indicating equal flux density values.  See text for more details on the analysis.}
\label{fig:WMAP_versus_WMAP_selyr1}
%\end{figure}
%\begin{figure}
\includegraphics[scale=0.6,angle=0]{figures/bias_WMAPyr1_versus_WMAPyr2345_LOG.epsi}
\caption{Comparison of the mean \textit{WMAP} 33-GHz flux density in years 2 to 5 with the mean \textit{WMAP} 33-GHz flux density in year 1 in three groups of sources, a low-, medium- and high-flux-density group, with a line indicating equal flux density values. The error bars are standard errors of the means. Calculations were performed in log space.}
\label{fig:bias_WMAPyr1_versus_WMAPyr2345}
\end{figure}

\subsubsection{Eddington bias in \textit{WMAP} 5-year data}

We have used the \textit{WMAP} 5-year, 33-GHz data to check whether the above effect is present in the \textit{WMAP} data.  We ran a blind search with a $4\sigma$ cut on the first-year map in the region $20\degr \leq b \leq 90\degr$; we excluded sources in the region $0\degr \leq b < 20\degr$ to avoid galactic contamination.  We followed this by a non-blind search on the five single-year maps, using prior knowledge of source positions from the first-year map.  We were then left with a list of sources selected in the first-year map, with flux densities in each separate year.  We corrected flux densities from the first-year map for the Eddington bias.  The correction for this bias becomes unreliable at low SNRs \citep[see][]{herranz}.  For this sample of sources, the correction starts becoming unreliable at SNR~$< 4.1$.  Having rerun the source-finding on the first-year map in the non-blind approach, some of the first-year flux densities were found to be significantly below $4\sigma$.  We, therefore, removed sources detected below $4.1\sigma$ in the first-year map.

In Fig.~\ref{fig:WMAP_versus_WMAP_selyr1}, we compare flux densities in year 1 with mean flux densities in years 2 to 5. We followed the same statistical analysis as described above, the results of which are shown in fig.~\ref{fig:bias_WMAPyr1_versus_WMAPyr2345}. As expected, in the low-flux-density group, the mean flux density in year 1, in which the sources were selected, is significantly higher, the effect being significant at a level of 3.5 $\sigma$. There is no apparent bias in the medium- and high-flux-density groups.

In order to verify that these results are not simply due to changes in telescope systematics over the course of the 5-year survey, we repeated the analysis selecting sources in the fifth-year map instead. This time, we selected sources from the whole sky. We removed sources detected below $4.1\sigma$ in the fifth-year map as well as sources in the strip $\vert b \vert \leq 20\degr$. Fig.~\ref{fig:WMAP_versus_WMAP_selyr5} compares flux densities in year 5 with mean flux densities in years 1 to 4 and Fig.~\ref{fig:bias_WMAPyr5_versus_WMAPyr1234} shows the results of our statistical analysis. We note that the error bars in Fig.~\ref{fig:bias_WMAPyr1_versus_WMAPyr2345} are bigger than those in \ref{fig:bias_WMAPyr5_versus_WMAPyr1234}.  This is simply because the analysis was carried out using data obtained from sky areas of different size.

In the low-flux-density group, the mean flux density in year 5, in which the sources were selected, is significantly higher, the effect being significant at a level of 6.9 $\sigma$. Again, there is no apparent bias in the medium- and high-flux-density groups.  We conclude that the bias is clearly present in the \textit{WMAP} 5-year data and that it is not related to the noise or to telescope systematics, but rather is a consequence of source variability.  In any survey, there will be preferential selection of those sources which were above their mean flux density values at the time of observation rather than below. The magnitude of the bias will not just depend on the general level of variability and the slope of the source counts, but also the survey duration and the typical variability timescale.  It should gradually decrease as the survey duration (i.e. making repeated measurements of the source flux density over a period of time) becomes comparable to the typical variability timescale.

\begin{figure}
\includegraphics[scale=0.6,angle=0]{figures/WMAP_versus_WMAP_selyr5.epsi}
\caption{Comparison of flux densities in year 5 with flux densities in years 1 to 4 for \textit{WMAP} 33-GHz data.  Sources were selected in the fifth-year map, with a line indicating equal flux density values.  See text for more details on the analysis.}
\label{fig:WMAP_versus_WMAP_selyr5}
%\end{figure} 
%\begin{figure}
\includegraphics[scale=0.6,angle=0]{figures/bias_WMAPyr5_versus_WMAPyr1234_LOG.epsi}
\caption{Comparison of the mean \textit{WMAP} 33-GHz flux densities in years 1 to 4 with the mean \textit{WMAP} 33-GHz flux densities in year 5 in three groups of sources, a low-, medium- and high-flux-density group, with a line indicating equal flux density values. The error bars are standard errors of the means. Calculations were performed in log space.}
\label{fig:bias_WMAPyr5_versus_WMAPyr1234}
\end{figure}

%%%%%%%%%%%%%%%%%%%%%%%%%%%%%%%%%%%%%%%%%%%%%%%%%%%%%%%%%%%%%%%%%%%
\section{Conclusions}\label{Conclusions}
%%%%%%%%%%%%%%%%%%%%%%%%%%%%%%%%%%%%%%%%%%%%%%%%%%%%%%%%%%%%%%%%%%%

In order to tie down the cm-wave spectra of sources at the high-flux-density end of the source population at cm wavelengths, we have made observations with the AMI-SA (13.9--18.2~GHz) and the VSA (33~GHz) of a complete sample of sources found with \textit{WMAP} at 33\:GHz and with $\mathrm{S_{33\:GHz} \geq 1.1\:Jy}$; AMI and VSA observations were scheduled so that variability is not an issue in the spectral measurements. We found that:
\begin{enumerate}

\item[(1)] the spectra are very different from those of bright sources at low frequency: 93~per~cent have spectra with $\alpha_{13.9}^{33.75} < 0.5$;

\item[(2)] 44 per cent have rising spectra ($\mathrm{\alpha_{13.9}^{33.75} < 0.0}$);

\item[(3)] in the group of 33 sources with average \textit{WMAP} and VSA flux densities $\mathrm{>2.26\:Jy}$, the average flux densities measured by \textit{WMAP} and VSA are fully consistent with each other. However, in the group of 32 sources with average \textit{WMAP} and VSA flux densities $\mathrm{<1.36\:Jy}$, the mean VSA flux density is lower than the mean \textit{WMAP} flux density, the discrepancy between the two being significant at a level of 4.9$\sigma$. We ascribe this to Eddington bias arising from variability.

\end{enumerate}

%%%%%%%%%%%%%%%%%%%%%%%%%%%%%%%%%%%%%%%%%%%%%%%%%%%%%%%%%%%%%%%%%%%
\section*{Acknowledgments}\label{acknowledgements}
%%%%%%%%%%%%%%%%%%%%%%%%%%%%%%%%%%%%%%%%%%%%%%%%%%%%%%%%%%%%%%%%%%%

We are grateful to the staff of the Mullard Radio Astronomy Observatory for the maintenance and operation of the AMI and to the staff of the Teide Observatory, Mullard Radio Astronomy Observatory and Jodrell Bank Observatory for assistance in the day-to-day operation of the VSA.  We are also grateful to the University of Cambridge and PPARC/STFC for funding and supporting the AMI.  We thank PPARC for funding and supporting the VSA project and the Spanish Ministry of Science and Technology for partial financial support (project AYA2001-1657).  MLD, TMOF, NHW, MO, CRG and TWS are grateful for support from PPARC/STFC studentships.

\begin{table*}
 \begin{minipage}{200mm}
 \caption{Results for individual sources.  The columns are the source name, taken from \newline the NEWPS catalogue; the Equatorial coordinates (J2000), from the PMN or GB6 \newline catalogues; the number of observations, $n$; the mean flux density, $\bar{S}$; and the spectral \newline index, $\alpha_{13.9}^{33.75}$.}
 \label{tab:sources}
 \begin{tabular}{@{}l c c c c r r r@{} c@{} l}
 \hline
 Source & $\mathrm{\alpha (J2000)}$ & $\mathrm{\delta (J2000)}$ & \multicolumn{2}{|c|}{$n$} & \multicolumn{2}{|c|}{${\bar{S}}$/Jy} & \multicolumn{3}{|c|}{$\alpha_{13.9}^{33.75}$}\\
  &  &  & AMI & VSA & AMI & VSA & & &\\   
 \hline
 J0029+0554  &  00 29 45.9  &$+$05 54 41&  3  &  3  &  0.65 &  0.66  & $   0.22~$&$\pm$&~0.19\\
 J0057+3021  &  00 57 48.3  &$+$30 21 14&  3  &  5  &  0.79 &  0.96  & $   0.06~$&$\pm$&~0.16\\
 J0105+4819  &  01 05 50.8  &$+$48 19 01&  5  &  6  &  0.49 &  0.58  & $   0.47~$&$\pm$&~0.16\\
 J0108+0134  &  01 08 38.7  &$+$01 34 51&  4  &  6  &  1.50 &  1.37  & $  -0.05~$&$\pm$&~0.12\\
 J0108+1319  &  01 08 52.7  &$+$13 19 17&  3  &  5  &  1.60 &  1.00  & $       ~$&$-  $&~    \\
 J0136+4751  &  01 36 58.8  &$+$47 51 27&  4  &  4  &  3.41 &  3.69  & $   0.04~$&$\pm$&~0.07\\
 J0152+2206  &  01 52 17.8  &$+$22 06 58&  3  &  4  &  0.88 &  0.95  & $   0.12~$&$\pm$&~0.19\\
 J0204+1514  &  02 04 50.8  &$+$15 14 10&  4  &  5  &  1.08 &  0.96  & $   0.33~$&$\pm$&~0.17\\
 J0205+3212  &  02 05 04.1  &$+$32 12 29&  4  &  4  &  2.76 &  2.98  & $  -0.05~$&$\pm$&~0.08\\
 J0221+3556  &  02 21 05.8  &$+$35 56 13&  4  &  5  &  1.27 &  1.13  & $   0.28~$&$\pm$&~0.14\\
 J0223+4259  &  02 23 14.5  &$+$42 59 19&  3  &  5  &  0.94 &  1.19  & $  -0.06~$&$\pm$&~0.20\\
 J0237+2848  &  02 37 52.4  &$+$28 48 14&  4  &  4  &  2.91 &  3.06  & $  -0.14~$&$\pm$&~0.14\\
 J0238+1637  &  02 38 38.5  &$+$16 37 04&  5  &  4  &  2.33 &  3.32  & $  -0.28~$&$\pm$&~0.05\\
 J0303+4716  &  03 03 34.8  &$+$47 16 19&  4  &  4  &  0.79 &  0.77  & $   0.00~$&$\pm$&~0.23\\
 J0319+4130  &  03 19 47.1  &$+$41 30 42&  4  &  4  & 14.93 & 11.51  & $   0.34~$&$\pm$&~0.06\\
 J0336+3218  &  03 36 30.0  &$+$32 18 36&  4  &  5  &  0.92 &  1.01  & $   0.07~$&$\pm$&~0.15\\
 J0339$-$0146  &  03 39 30.4  &$-$01 46 38&  3  &  5  &  1.92 &  1.89  & $   0.10~$&$\pm$&~0.13\\
 J0418+3801  &  04 18 22.4  &$+$38 01 47&  4  &  5  &  5.41 &  7.20  & $  -0.35~$&$\pm$&~0.06\\
 J0423$-$0120  &  04 23 15.8  &$-$01 20 34&  3  &  5  &  4.44 &  4.85  & $  -0.01~$&$\pm$&~0.08\\
 J0423+4150  &  04 23 55.7  &$+$41 50 06&  4  &  5  &  1.32 &  1.25  & $   0.32~$&$\pm$&~0.13\\
 J0424+0036  &  04 24 46.6  &$+$00 36 05&  1  &  1  &  0.50 &  0.40  & $   0.58~$&$\pm$&~0.26\\
 J0433+0521  &  04 33 11.0  &$+$05 21 13&  3  &  5  &  2.95 &  2.82  & $   0.33~$&$\pm$&~0.09\\
 J0437+2940  &  04 37 04.7  &$+$29 40 02&  4  &  4  &  4.47 &  2.45  & $   0.94~$&$\pm$&~0.15\\
 J0449+1121  &  04 49 07.6  &$+$11 21 25&  2  &  4  &  0.86 &  1.02  & $   0.55~$&$\pm$&~0.28\\
 J0501$-$0159  &  05 01 12.9  &$-$01 59 21&  3  &  5  &  0.92 &  0.99  & $   0.16~$&$\pm$&~0.17\\
 J0533+4822  &  05 33 15.6  &$+$48 22 59&  3  &  4  &  1.00 &  1.08  & $  -0.13~$&$\pm$&~0.16\\
 J0555+3948  &  05 55 31.7  &$+$39 48 45&  3  &  4  &  3.13 &  2.74  & $   0.41~$&$\pm$&~0.13\\
 J0646+4451  &  06 46 31.4  &$+$44 51 22&  3  &  4  &  3.20 &  2.76  & $   0.21~$&$\pm$&~0.14\\
 J0733+5022  &  07 33 52.8  &$+$50 22 18&  4  &  5  &  0.78 &  0.80  & $   0.00~$&$\pm$&~0.14\\
 J0738+1742  &  07 38 07.6  &$+$17 42 26&  4  &  4  &  0.78 &  0.80  & $   0.07~$&$\pm$&~0.16\\
 J0739+0137  &  07 39 18.2  &$+$01 37 06&  3  &  5  &  1.26 &  1.80  & $   0.09~$&$\pm$&~0.15\\
 J0750+1231  &  07 50 51.2  &$+$12 31 13&  3  &  4  &  4.18 &  4.02  & $  -0.01~$&$\pm$&~0.08\\
 J0757+0956  &  07 57 06.4  &$+$09 56 21&  3  &  4  &  1.28 &  1.65  & $  -0.19~$&$\pm$&~0.13\\
 J0825+0309  &  08 25 49.5  &$+$03 09 25&  4  &  5  &  1.30 &  1.89  & $  -0.24~$&$\pm$&~0.13\\
 J0830+2410  &  08 30 52.3  &$+$24 10 47&  4  &  3  &  1.11 &  1.28  & $  -0.09~$&$\pm$&~0.16\\
 J0840+1312  &  08 40 48.0  &$+$13 12 37&  4  &  3  &  0.97 &  1.06  & $   0.21~$&$\pm$&~0.19\\
 J0854+2006  &  08 54 48.4  &$+$20 06 47&  3  &  3  &  2.85 &  3.37  & $  -0.30~$&$\pm$&~0.09\\
 J0909+0121  &  09 09 09.5  &$+$01 21 38&  3  &  5  &  1.21 &  1.32  & $   0.08~$&$\pm$&~0.14\\
 J0920+4441  &  09 20 58.7  &$+$44 41 44&  3  &  4  &  1.92 &  2.34  & $  -0.32~$&$\pm$&~0.12\\
 J0927+3902  &  09 27 03.0  &$+$39 02 18&  4  &  5  &  9.75 &  8.49  & $   0.19~$&$\pm$&~0.05\\
 J0948+4039  &  09 48 55.2  &$+$40 39 56&  3  &  4  &  1.59 &  1.55  & $  -0.07~$&$\pm$&~0.15\\
 J0958+4725  &  09 58 19.9  &$+$47 25 14&  5  &  6  &  1.26 &  1.16  & $   0.23~$&$\pm$&~0.11\\
 J1033+4115  &  10 33 03.9  &$+$41 15 59&  2  &  3  &  0.81 &  0.96  & $  -0.22~$&$\pm$&~0.18\\
 J1038+0512  &  10 38 47.7  &$+$05 12 16&  4  &  6  &  1.41 &  1.21  & $   0.40~$&$\pm$&~0.13\\
 J1041+0610  &  10 41 17.6  &$+$06 10 02&  3  &  7  &  1.38 &  1.36  & $   0.07~$&$\pm$&~0.15\\
 J1058+0133  &  10 58 30.5  &$+$01 33 46&  3  &  6  &  4.59 &  5.92  & $  -0.18~$&$\pm$&~0.06\\
 J1130+3815  &  11 30 54.6  &$+$38 15 10&  5  &  6  &  1.35 &  1.40  & $  -0.02~$&$\pm$&~0.10\\
 J1153+4931  &  11 53 24.7  &$+$49 31 13&  5  &  6  &  1.20 &  1.29  & $   0.01~$&$\pm$&~0.11\\
 J1159+2914  &  11 59 32.1  &$+$29 14 53&  4  &  5  &  1.56 &  1.61  & $  -0.20~$&$\pm$&~0.13\\
 J1219+0549  &  12 19 18.0  &$+$05 49 39&  0  &  4  &  $-$~~&  1.86  & $   0.80~$&$\pm$&~0.35\\
 J1229+0203  &  12 29 05.6  &$+$02 03 09&  3  &  5  & 29.20 & 26.98  & $   0.07~$&$\pm$&~0.05\\
 J1230+1223  &  12 30 48.8  &$+$12 23 36&  3  &  5  & 24.49 & 17.67  & $   0.41~$&$\pm$&~0.05\\
 J1310+3220  &  13 10 29.5  &$+$32 20 51&  4  &  5  &  1.21 &  1.79  & $  -0.61~$&$\pm$&~0.11\\
 J1331+3030  &  13 31 08.0  &$+$30 30 35&  4  &  5  &  3.34 &  2.06  & $   0.63~$&$\pm$&~0.10\\
 J1347+1217  &  13 47 33.4  &$+$12 17 17&  3  &  4  &  1.36 &  1.03  & $   0.62~$&$\pm$&~0.20\\
 J1357+1919  &  13 57 04.1  &$+$19 19 19&  3  &  4  &  1.98 &  2.36  & $  -0.35~$&$\pm$&~0.13\\
 J1419+3822  &  14 19 45.9  &$+$38 22 01&  4  &  4  &  0.60 &  0.86  & $  -0.08~$&$\pm$&~0.18\\
 J1504+1029  &  15 04 24.0  &$+$10 29 43&  3  &  4  &  1.60 &  1.50  & $   0.00~$&$\pm$&~0.17\\
 J1516+0014  &  15 16 40.7  &$+$00 14 57&  3  &  4  &  1.01 &  1.07  & $   0.06~$&$\pm$&~0.22\\
 J1549+0237  &  15 49 30.0  &$+$02 37 01&  3  &  4  &  2.16 &  2.19  & $   0.08~$&$\pm$&~0.15\\
 J1550+0527  &  15 50 35.2  &$+$05 27 06&  3  &  4  &  2.82 &  2.84  & $   0.02~$&$\pm$&~0.10\\
 \end{tabular}							
\end{minipage}							
\end{table*}

\begin{table*}
\begin{minipage}{200mm}
 \contcaption{}
 \begin{tabular}{@{}l c c c c r r r@{} c@{} l}
 J1608+1029  &  16 08 46.4  &$+$10 29 05&  3  &  4  &  1.34 &  1.44  & $   0.11~$&$\pm$&~0.19\\
 J1613+3412  &  16 13 40.8  &$+$34 12 41&  4  &  4  &  3.21 &  2.68  & $   0.23~$&$\pm$&~0.08\\
 J1635+3808  &  16 35 15.6  &$+$38 08 13&  4  &  4  &  2.48 &  2.58  & $  -0.12~$&$\pm$&~0.12\\
 J1638+5720  &  16 38 13.0  &$+$57 20 29&  4  &  4  &  1.92 &  2.45  & $  -0.34~$&$\pm$&~0.12\\
 J1642+3948  &  16 42 58.0  &$+$39 48 42&  4  &  4  &  5.30 &  5.26  & $   0.18~$&$\pm$&~0.07\\
 J1651+0459  &  16 51 09.2  &$+$04 59 33&  3  &  4  &  2.63 &  1.33  & $   1.03~$&$\pm$&~0.18\\
 J1720$-$0058  &  17 20 29.7  &$-$00 58 37&  0  &  5  &$-$~~&  3.87  & $        $&$ - $&~    \\
 J1727+4530  &  17 27 28.4  &$+$45 30 49&  4  &  6  &  0.57 &  0.94  & $  -0.67~$&$\pm$&~0.17\\
 J1734+3857  &  17 34 20.5  &$+$38 57 45&  4  &  4  &  0.89 &  0.93  & $  -0.21~$&$\pm$&~0.17\\
 J1740+5211  &  17 40 36.6  &$+$52 11 47&  4  &  4  &  1.02 &  0.95  & $   0.12~$&$\pm$&~0.17\\
 J1743$-$0350  &  17 43 59.2  &$-$03 50 06&  4  &  5  &  3.80 &  3.66  & $   0.28~$&$\pm$&~0.07\\
 J1751+0938  &  17 51 32.7  &$+$09 38 58&  3  &  4  &  5.08 &  6.36  & $  -0.19~$&$\pm$&~0.07\\
 J1753+2847  &  17 53 42.5  &$+$28 47 58&  3  &  4  &  1.78 &  1.85  & $  -0.14~$&$\pm$&~0.13\\
 J1801+4404  &  18 01 32.2  &$+$44 04 09&  4  &  4  &  1.40 &  1.47  & $   0.04~$&$\pm$&~0.15\\
 J1824+5650  &  18 24 06.8  &$+$56 50 59&  4  &  5  &  1.28 &  1.34  & $  -0.26~$&$\pm$&~0.12\\
 J1829+4844  &  18 29 32.1  &$+$48 44 46&  3  &  4  &  2.70 &  2.32  & $   0.16~$&$\pm$&~0.13\\
 J1955+5131  &  19 55 42.3  &$+$51 31 54&  4  &  3  &  1.18 &  1.50  & $  -0.63~$&$\pm$&~0.14\\
 J1959+4034  &  19 59 21.8  &$+$40 34 28&  0  &  3  &  $-$~~& 37.43  & $        $&$ - $&~    \\
 J2123+0535  &  21 23 43.4  &$+$05 35 14&  3  &  3  &  1.41 &  1.41  & $   0.00~$&$\pm$&~0.19\\
 J2134$-$0153  &  21 34 10.4  &$-$01 53 25&  3  &  2  &  1.90 &  1.75  & $   0.27~$&$\pm$&~0.19\\
 J2136+0041  &  21 36 38.6  &$+$00 41 54&  3  &  6  &  5.76 &  4.22  & $   0.42~$&$\pm$&~0.05\\
 J2139+1423  &  21 39 01.5  &$+$14 23 37&  3  &  4  &  2.13 &  1.75  & $   0.33~$&$\pm$&~0.08\\
 J2143+1743  &  21 43 35.6  &$+$17 43 54&  4  &  6  &  0.50 &  0.57  & $  -0.23~$&$\pm$&~0.14\\
 J2148+0657  &  21 48 05.5  &$+$06 57 36&  3  &  5  &  5.52 &  5.07  & $   0.19~$&$\pm$&~0.05\\
 J2202+4216  &  22 02 44.3  &$+$42 16 39&  4  &  5  &  3.47 &  4.33  & $  -0.26~$&$\pm$&~0.05\\
 J2203+1725  &  22 03 26.7  &$+$17 25 42&  4  &  5  &  1.18 &  1.25  & $  -0.07~$&$\pm$&~0.11\\
 J2203+3145  &  22 03 15.8  &$+$31 45 38&  4  &  5  &  2.62 &  2.60  & $  -0.08~$&$\pm$&~0.08\\
 J2212+2355  &  22 12 05.9  &$+$23 55 31&  4  &  4  &  0.95 &  0.96  & $  -0.14~$&$\pm$&~0.12\\
 J2218$-$0335  &  22 18 51.8  &$-$03 35 40&  3  &  4  &  1.65 &  1.33  & $   0.47~$&$\pm$&~0.15\\
 J2225+2118  &  22 25 37.6  &$+$21 18 17&  4  &  4  &  1.11 &  1.27  & $  -0.07~$&$\pm$&~0.11\\
 J2232+1143  &  22 32 36.6  &$+$11 43 54&  4  &  4  &  3.79 &  4.52  & $  -0.29~$&$\pm$&~0.05\\
 J2236+2828  &  22 36 20.8  &$+$28 28 56&  5  &  5  &  0.99 &  1.60  & $  -0.51~$&$\pm$&~0.10\\
 J2253+1608  &  22 53 58.0  &$+$16 08 53&  4  &  5  &  7.17 & 10.73  & $  -0.56~$&$\pm$&~0.05\\
 J2327+0940  &  23 27 33.1  &$+$09 40 02&  4  &  6  &  1.81 &  2.18  & $  -0.27~$&$\pm$&~0.08\\
 J2335$-$0131  &  23 35 20.1  &$-$01 31 14&  3  &  6  &  0.56 &  0.79  & $   0.21~$&$\pm$&~0.17\\
 J2354+4553  &  23 54 21.9  &$+$45 53 00&  4  &  5  &  0.98 &  0.77  & $   0.32~$&$\pm$&~0.14\\
  \hline
 \end{tabular}
\end{minipage}
\end{table*}

\label{lastpage}

\end{document}